\documentclass[12pt]{article}
\pdfoutput=1
\usepackage{graphicx}
\usepackage{subfigure}
\usepackage{epstopdf}
\usepackage{cite}
\usepackage{xspace}
\usepackage{amsmath,amssymb,textcomp}
\usepackage{booktabs}

\newcommand\pubnumber{}
\newcommand\pubdate{\today}

\def\oxford{University of Oxford, Denys Wilkinson Building, Keble Road, \\ OX1 3RH, United Kingdom}

\def\Title#1{\begin{center} {\Large #1 } \end{center}}
\def\Author#1{\begin{center}{ \sc #1} \end{center}}
\def\Address#1{\begin{center}{ \it #1} \end{center}}

\newcommand\pubblock{\rightline{\begin{tabular}{l} \pubnumber \\ \pubdate \end{tabular}}}
\newenvironment{Abstract}{\begin{quotation}  }{\end{quotation}}

\def\beq{\begin{equation}}
\def\eeq#1{\label{#1}\end{equation}}
\def\eeqn{\end{equation}}

\def\beqa{\begin{eqnarray}}
\def\eeqa#1{\label{#1}\end{eqnarray}}
\def\eeqan{\end{eqnarray}}

\let\bar=\overbar

\def\Dslash{\not{\hbox{\kern-4pt $D$}}}
\def\dslash{\not{\hbox{\kern-2pt $\del$}}}

\def\msb{{\bar{\ssstyle M \kern -1pt S}}}

\newcommand{\CP}{\ensuremath{C\!P}\xspace}

\newcommand{\PKp}{\ensuremath{K^+}\xspace}
\newcommand{\PKm}{\ensuremath{K^-}\xspace}
\newcommand{\PPip}{\ensuremath{\pi^+}\xspace}
\newcommand{\PPim}{\ensuremath{\pi^-}\xspace}

\newcommand{\KsKK}{\ensuremath{K^0_{\rm S} K^+K^-}\xspace}
\newcommand{\KsPiPi}{\ensuremath{K^0_{\rm S}\pi^+\pi^-}\xspace}
\newcommand{\KsHH}{\ensuremath{K^0_{\rm S} h^+h^-}\xspace}

\newcommand{\PD}{\ensuremath{D}\xspace}
\newcommand{\PDb}{\ensuremath{\kern 0.2em\overline{\kern -0.2em D}}\xspace}
\newcommand{\PDz}{\ensuremath{D^0}\xspace}
\newcommand{\PDzb}{\ensuremath{\kern 0.2em\overline{\kern -0.2em D}{}^0}\xspace}

\newcommand{\xD}{\ensuremath{x_D}\xspace}
\newcommand{\yD}{\ensuremath{y_D}\xspace}

\newcommand{\mxsq}{\ensuremath{m_{12}^2}\xspace}
\newcommand{\mysq}{\ensuremath{m_{13}^2}\xspace}
\newcommand{\axy}{\ensuremath{a_{12,13}}\xspace}
\newcommand{\ayx}{\ensuremath{a_{13,12}}\xspace}
\newcommand{\dxy}{\ensuremath{\delta_{12,13}}\xspace}
\newcommand{\dyx}{\ensuremath{\delta_{13,12}}\xspace}

\newcommand{\rCP}{\ensuremath{r_{CP}}\xspace}
\newcommand{\aCP}{\ensuremath{\alpha_{CP}}\xspace}

\newcommand{\ci}{\ensuremath{c_i}\xspace}
\newcommand{\si}{\ensuremath{s_i}\xspace}
\newcommand{\Ti}{\ensuremath{T_i}\xspace}
\newcommand{\Tmi}{\ensuremath{T_{-i}}\xspace}

\newcommand{\dg}{\ensuremath{^{\circ}}\xspace}

\newcommand\T{\rule{0pt}{2.6ex}}
\newcommand\B{\rule[-1.2ex]{0pt}{0pt}}

\begin{document}
\begin{titlepage}
\pubblock

\vfill
\Title{Model-independent \PDz--\PDzb mixing and \CP violation \\ \vspace*{0.1cm}  studies  with $\PDz\to\KsPiPi$ and 
$\PDz\to\KsKK$}
\vfill
\Author{C. Thomas and G. Wilkinson}
\Address{\oxford}
\vfill
\begin{Abstract}
\noindent
Simulation studies are performed to assess the sensitivity of a model-independent analysis of the flavour-tagged decays $\PDz\to\KsPiPi$
and $\PDz\to\KsKK$ to mixing and \CP violation. The analysis takes as input
measurements of the $D$ decay strong-phase parameters that are accessible in quantum-correlated \PD--\PDb pairs produced at the $\psi (3770)$ resonance.   It is shown that the model-independent approach is 
 well suited to the very large data sets expected at an upgraded LHCb experiment, or future high  luminosity $e^+e^-$ facility, and that with 100M \KsPiPi decays a statistical precision of around $0.01$ and $0.7^\circ$ is achievable on the \CP violation parameters \rCP and \aCP, respectively.  Even with this very large sample the systematic uncertainties associated with the strong-phase parameters will not be limiting, assuming that full use is made of the available $\psi(3770)$ data sets of CLEO-c and BES-III.
Furthermore, it is demonstrated that  large flavour-tagged samples can themselves be exploited to provide information on the strong-phase parameters,  a feature that will be beneficial in the measurement of the CKM angle $\gamma/\phi_3$ with $B^- \to DK^-$ decays.
\end{Abstract}
\vfill
\vfill
\end{titlepage}
\def\thefootnote{\fnsymbol{footnote}}
\setcounter{footnote}{0}

%%%%%%%%%%%%%%%%%%%%%%%%%%%%%%%%%%%%%%%%%%%%%%%%%%%%%%%%%%%%%%%%%%%%%%%%%
%%
%%   use this format to include an .eps figure into your paper
%%
%\begin{figure}[htb]
%\centering
%\includegraphics[height=1.5in]{magnet}
%\caption{Plan of the magnet used in the mesmeric studies.}
%\label{fig:magnet}
%\end{figure}
%%%%%%%%%%%%%%%%%%%%%%%%%%%%%%%%%%%%%%%%%%%%%%%%%%%%%%%%%%%%%%%%%%%%%%%%%%%
%\Acknowledgements
%I am grateful to Don Alfonso d'Alba for certain services essential to 
%this investigation.

\section{Introduction}
\label{sec:intro}

In the last few years measurements have been performed at the $B$-factories and
the Tevatron which, when taken together, reveal the presence
of mixing in the \PDz--\PDzb system~\cite{BELLEMIX,BaBarMIX,CDFMIX,MIXKSPP_BELLE,MIXKSPP_BaBar,BELLEKSKK}. Improved sensitivity 
to this phenomenon will come from the LHCb experiment, which has already 
published a first study with early data~\cite{LHCBYCP}. Interest in the charm sector has been raised still further with the recent announcement of evidence for \CP violation in time-integrated $\PDz\to\PKp\PKm$ and $\PDz\to\PPip\PPim$  decays\footnote{Unless stated otherwise, the charge-conjugate mode is implicit throughout this article.}~\cite{LHCBDACP,CDFDACP}. This effect, if confirmed by future measurements, is almost certainly attributable to {\it  direct} \CP violation~\cite{HFAG}.  A natural next step in the experimental programme is  to intensify the search for {\it indirect} \CP violation, which manifests itself through mixing-related observables. 
 In the Standard Model (SM) such \CP violation is expected to be very small and so any non-zero effect would be indicative of possible New Physics (NP) contributions~\cite{BIGI}. 

A particularly powerful decay mode for flavour-tagged time-dependent \PDz--\PDzb studies is the channel $\PDz\to\KsPiPi$, where the interference between intermediate resonances provides sensitivity to both the magnitude and sign of the mixing parameters, and allows for probes of indirect \CP violation.  The analysis of this mode was pioneered by CLEO~\cite{MIXKSPP_CLEO}, and higher precision results have been published by Belle~\cite{MIXKSPP_BELLE} and BaBar~\cite{MIXKSPP_BaBar}, based on samples of around $0.5\,$M and $0.7\,$M events respectively. 
Studies based on larger data sets can soon be expected from LHCb.   In time, sample sizes of ${\cal O}(10^{8})$ events may become available from future experiments, such as Belle-II~\cite{BELLEII}, SuperB~\cite{SUPERB} and the LHCb upgrade~\cite{LHCBUPGRADE,LHCBUPGRADE2}.  It is important, therefore, to understand what sensitivity to \CP violation can be attained with large data sets of $\PDz\to\KsPiPi$ decays, and what challenges such a measurement will be confronted with.

The  $\PDz\to\KsPiPi$ analysis is performed in the Dalitz space of the three-body final state. Measurement of the mixing and \CP violation parameters requires a good understanding of the decay structure within this Dalitz space. The established method is to develop an amplitude model of the \PDz decay on flavour-tagged data, and then exploit this model in an unbinned time-dependent likelihood fit.  The drawback to this approach is that the results have a  systematic uncertainty that is associated with the assumptions of the model.  The size of the uncertainty is difficult to assess, and the degree to which it can be decreased with larger data samples is unknown.   An alternative strategy, first proposed in~\cite{BPV}, is to perform a binned, model-independent analysis, where the necessary inputs on the \PDz decay come from auxiliary measurements performed on quantum-correlated \PD--\PDb decays at the $\psi(3770)$ resonance.  In this case the systematic uncertainties associated with the model are replaced with measurement uncertainties coming from these external inputs.  No data analysis has yet been presented which follows this model-independent approach.

Identical considerations apply to the  channel $\PDz\to\KsKK$. Although suppressed by a factor of around six compared to $\PDz\to\KsPiPi$, this decay can be exploited in an identical manner for mixing and \CP violation studies.   A model-dependent analysis has been performed by BaBar~\cite{MIXKSPP_BaBar}, and Belle have studied the time-evolution of the \CP-odd component of the decay~\cite{BELLEKSKK}.   The model-independent study of the full Dalitz space is also an appealing analysis option for this decay mode, but one that has not yet been pursued.

This paper presents the results of simulation studies performed to investigate the sensitivity of the model-independent approach to the decays  $\PDz\to\KsPiPi$  and $\PDz\to\KsKK$.   The statistical precision of a binned analysis is assessed and compared with that  of a model-dependent, unbinned study.  
The sensitivity is calculated both for the sample sizes that currently exist, and for those which will become available over the coming years.  
When considering future large data sets, the point at which the analysis will become limited by the existing measurements of the  $D$ meson decay properties from the $\psi(3770)$, available from the CLEO-c experiment~\cite{CLEOCISI}, is determined, together with what improvement in the knowledge of these properties will be required.  
%When considering future, large data sets it is asked  at what point the analysis will become  limited by  the existing measurements of the  $D$ meson decay properties from the $\psi(3770)$, available from the CLEO-c experiment~\cite{CLEOCISI}, and what improvement in the knowledge of these properties will be required. 
Various fit strategies are explored which make different use of external inputs.  It is shown that other useful results can be obtained from the analysis, aside from the mixing and \CP violation parameters.  

Model-independent studies of mixing and \CP violation with other \PDz decay modes are possible ($e.g.$ see Ref.~\cite{SNEHA}), all of which can benefit from measurements of strong-phase parameters and related properties in $\psi(3770)$ data~\cite{OTHERCLEO}. 

\section{Formalism}
\label{sec:formalism}

In the neutral \PD meson system the mass eigenstates, $D_{1,2}$, are related to the flavour eigenstates \PDz and \PDzb as follows:
\begin{equation}
|D_{1,2}\rangle = p|\PDz\rangle \pm q|\PDzb\rangle
\end{equation}
where the coefficients satisfy $|p|^2 + |q|^2 = 1$ and 
\begin{equation}
\rCP e^{i \aCP} \equiv \frac{q}{p}.
\label{eq:cpvdefinition}
\end{equation}
Indirect \CP violation occurs if $\rCP \ne 1$ and/or $\aCP \ne 0$.
Charm mixing is conventionally parameterised by the quantities \xD and \yD, defined as
\begin{equation}
\xD \equiv \frac{M_2 - M_1}{\Gamma},  \qquad \yD \equiv \frac{\Gamma_2 - \Gamma_1}{2\Gamma}
\label{eq:xydefinition}
\end{equation}
where $M_{1,2}$ and $\Gamma_{1,2}$ are the mass and width of the two neutral \PD meson mass eigenstates, and $\Gamma$ the mean decay width of the mass eigenstates.  
A global  analysis of all current measurements~\cite{HFAG} yields the results $\xD = (0.63^{+0.19}_{-0.20}) \%$, $\yD = (0.75 \pm 0.12) \%$, $\rCP = 0.89^{+0.17}_{-0.15}$ and $\aCP = (-10.1^{+9.4}_{-8.8})\dg$.  The values of the latter two parameters are consistent with the null \CP violation hypothesis.

The time evolution of neutral \PDz mesons proceeds as follows:
\begin{align}
|\PDz(t)\rangle  & = \phantom{\frac{p}{q}}\,g_+(t)|\PDz\rangle + \frac{q}{p}         \,g_-(t)|\PDzb\rangle, \\
|\PDzb(t)\rangle & = \frac{p}{q}          \,g_-(t)|\PDz\rangle + \phantom{\frac{q}{p}}\,g_+(t)|\PDzb\rangle,
\end{align}
where
\begin{equation}
g_\pm(t) \equiv \frac{1}{2}\Big(e^{-i(M_1 - i\Gamma_1/2)t} \pm e^{-i(M_2 - i\Gamma_2/2)t}\Big).
\end{equation}

%Given the small values of the mixing parameters, the functions $g_+(t)$ and $g_-(t)$ can be written:
%\begin{align}
%g_+(t) & \sim e^{-imt} e^{-\Gamma t/2}\left[1 + \frac{1}{8}(i\xD + \yD)^2 (\Gamma t)^2 + \dots\right], \\
%g_-(t) & \sim e^{-imt} e^{-\Gamma t/2}\left[\frac{1}{2}(i\xD + \yD)(\Gamma t) + \dots\right].
%\end{align}

%When squared, these quantities yield:
%\begin{align}
%\label{eq:gpsq} |g_+(t)|^2 & \sim e^{-\Gamma t} [1], \\
%\label{eq:gmsq} |g_-(t)|^2 & \sim e^{-\Gamma t} [(\yD^2 + \xD^2)(\Gamma t)^2/4], \\
%\label{eq:gpgm} g_+^*(t)g_-(t) & \sim e^{-\Gamma t}[(i\xD + \yD)(\Gamma t)/2], \\
%\label{eq:gmgp} g_-^*(t)g_+(t) & \sim e^{-\Gamma t}[(-i\xD + \yD)(\Gamma t)/2], 
%\end{align}
%where terms with order greater than $\mathcal{O}(\xD^2), \mathcal{O}(\xD\yD), \mathcal{O}(\yD^2)$ have been neglected.

Consideration is now given to the specific case in which the \PDz decays to the final state \KsPiPi (the formalism presented is identical for \KsKK).   The final-state particles are labeled as follows: 1 for $K^0_{\rm S}$,  2 for $\pi^+$ and 3 for $\pi^-$.
The Dalitz plot decay amplitude density of $\PDz\to\KsPiPi$ is defined as
\begin{equation}
\mathcal{A}_{\PDz}(\mxsq,\mysq) \equiv \axy e^{i\dxy},
\end{equation}
where \mxsq and \mysq specify coordinates on the $\PDz\to\KsPiPi$ Dalitz plot, \axy is the modulus of the amplitude and $\dxy$ is a \CP-conserving strong-phase.
% (the numbers label the final-state particles as follows: 1 is \PKs, 2 is \PPip and 3 is \PPim). 
The decay amplitude density is independent of the time evolution of the \PDz system.
Assuming there is no direct \CP violation in decays to the final state in question, the equivalent decay amplitude density for \PDzb decays is
\begin{equation}
\mathcal{A}_{\PDzb}(\mxsq,\mysq) = \mathcal{A}_{\PDz}(\mysq,\mxsq) \equiv \ayx e^{i\dyx}.
\end{equation}

The time-dependent probability density in which mixing and indirect \CP violation are allowed is
\begin{align}
{\cal {P}}_{\PDz}(\mxsq,\mysq,t) 
& = \Gamma \left|g_+(t)\mathcal{A}_{\PDz}(\mxsq,\mysq) + \frac{q}{p}g_-(t)\mathcal{A}_{\PDzb}(\mxsq,\mysq)\right|^2 \nonumber \\
& = \Gamma |g_+(t)|^2 \axy^2 + \Gamma \left|\frac{q}{p}\right|^2 |g_-(t)|^2 \ayx^2 \nonumber\\
& + \Gamma \, \mathfrak{Re}\left\{\frac{q}{p} g_+^*(t)g_-(t) \, \axy\,\ayx \, e^{i(-\dxy+\dyx)}\right\} \nonumber\\
& + \Gamma \, \mathfrak{Re}\left\{\left(\frac{q}{p}\right)^{\!*} g_+(t)g_-^*(t) \, \axy\,\ayx \, e^{i(\dxy-\dyx)}\right\},
\end{align}
which, ignoring terms of $\mathcal{O}(\xD^2)$, $\mathcal{O}(\xD\yD)$ and $\mathcal{O}(\yD^2)$, is equal to
\begin{align}
\label{eq:PDFxyt} {\cal{P}}_{\PDz}(\mxsq,\mysq,t) & = \Gamma e^{-\Gamma t} \Big[\axy^2 + \rCP\,\axy\,\ayx \, \Gamma t \, \big\{\yD \cos(\dxy - \dyx - \aCP) \nonumber \\
& \qquad\qquad + \xD \sin(\dxy - \dyx - \aCP)\big\}\Big].
%\label{eq:PDFxyt} {\cal{P}}_{\PDz}(\mxsq,\mysq,t) & = \Gamma e^{-\Gamma t} \cdot \axy^2 \nonumber\\
%& + \Gamma e^{-\Gamma t} \cdot \rCP\axy\ayx \cdot \frac{\Gamma t}{2} \cdot \mathfrak{Re}\big\{(i\xD + \yD) \, e^{i(-\dxy+\dyx - \aCP)} \big\} \nonumber\\ 
%& + \Gamma e^{-\Gamma t} \cdot \rCP\axy\ayx \cdot \frac{\Gamma t}{2} \cdot \mathfrak{Re}\big\{(-i\xD + \yD) \, e^{i(\dxy-\dyx + \aCP)} \big\} \nonumber \\
%& = \Gamma e^{-\Gamma t} \Big[\axy^2 + \rCP\axy\ayx \cdot \Gamma t \cdot \big\{\yD \cos(\dxy - \dyx - \aCP) \nonumber \\
%& + \xD \sin(\dxy - \dyx - \aCP)\big\}\Big].
\end{align}
Integrating the probability density between times $t_a$ and $t_b$ yields
\begin{align}
\int^{t_b}_{t_a} & {\cal{P}}_{\PDz}(\mxsq,\mysq,t) \, \text{d}t  = \nonumber \\
& (e^{-\Gamma t_a} - e^{-\Gamma t_b}) \axy^2 + \left[\Gamma(e^{-\Gamma t_a}t_a - e^{-\Gamma t_b}t_b) + (e^{-\Gamma t_a} - e^{-\Gamma t_b})\right] \nonumber \\
& \times \rCP\axy\,\ayx  \big\{ \yD [\cos(\dxy - \dyx)\cos(\aCP) + \sin(\dxy - \dyx)\sin(\aCP)] \nonumber \\
& + \xD [\sin(\dxy - \dyx)\cos(\aCP) - \cos(\dxy - \dyx)\sin(\aCP)]\big\}.
\label{eq:unbinned2}
\end{align}
The formalism for the time evolution of the \PDzb decay probability is identical, except \rCP is replaced by 1/\rCP and \aCP is replaced by $-\aCP$.

If a model is available to describe the variation of amplitude  across the Dalitz plot then Eqn.~\ref{eq:unbinned2} may be used in a maximum likelihood fit, unbinned in Dalitz space, to determine the mixing and \CP violation parameters in the analysis of a given data sample.  If however a model-independent approach is pursued then it is necessary to partition the Dalitz plane.   A total of $2N$ regions are defined, symmetric under the exchange $m^2_{12} \leftrightarrow m^2_{13}$ and labelled from $-N$ to $+N$ (excluding zero).  Then the population in bin $i$ is given by
\begin{align}
\label{eq:KpTBinCPV}
%K^\prime_i|_{t_a}^{t_b} 
\int _i \int^{t_b}_{t_a} & {\cal{P}}_{\PDz}(\mxsq,\mysq,t) \, \text{d}t \, \text{d}m_{12}^2 \, \text{d}m_{13}^2 = \nonumber \\
& n\Big\{ (e^{-\Gamma t_a} - e^{-\Gamma t_b}) \Ti + \left[\Gamma(e^{-\Gamma t_a}t_a - e^{-\Gamma t_b}t_b) + (e^{-\Gamma t_a} - e^{-\Gamma t_b})\right] \nonumber \\
& \times \big\{\rCP\sqrt{\Ti\Tmi} (\yD [\ci\cos(\aCP) + \si\sin(\aCP)] + \xD [\si\cos(\aCP) - \ci\sin(\aCP)])\big\} \Big\}
\end{align}
where $n$ is a normalisation factor and \Ti, \ci and \si are defined as follows:
\begin{align}
\Ti & \equiv \int_i \axy^2 \,\text{d}\mxsq\,\text{d}\mysq \,, \\
\ci & \equiv \frac{1}{\sqrt{\mathstrut\Ti\Tmi}}\int_i \axy\,\ayx \cos(\dxy - \dyx) \,\text{d}\mxsq\,\text{d}\mysq \,, \\
\si & \equiv \frac{1}{\sqrt{\mathstrut\Ti\Tmi}}\int_i \axy\,\ayx \sin(\dxy - \dyx) \,\text{d}\mxsq\,\text{d}\mysq \,.
\end{align}
These expressions make use of the relations $c_{-i} = c_i$ and $s_{-i} = -s_i$.  In the limit of no mixing or \CP violation \Ti is proportional to the population in bin $i$ coming from decays of $D$ mesons of a known flavour.   The quantity \ci (\si) is the cosine (sine) of the average strong-phase difference between \PDz and \PDzb decays within bin $i$, weighted by the decay rate.

As external measurements are available for \ci and \si (see Sect.~\ref{sec:cisimeasurements}),  and the values of \Ti may also be obtained from external data sets, or indeed left as free parameters in the fit (see Sect.~\ref{sec:cisiinput}), it is possible to use the observed population of the Dalitz and decay time bins to determine the mixing and \CP parameters independent of any model assumptions.

\section{Available and future measurements of $\boldsymbol{c_i}$ and  $\boldsymbol{s_i}$}
\label{sec:cisimeasurements}

The CLEO collaboration have published results which exploit the  quantum-correlated nature of $\psi(3770)\to\PD\PDb$ decays to measure the \ci and \si coefficients for both $\PDz\to\KsPiPi$ and $\PDz\to\KsKK$~\cite{CLEOCISI}.  
The study was motivated by the need to provide input for the model-independent measurement of the CKM unitarity angle $\gamma/\phi_3$ in $B^- \to\PD K^-$ decays~\cite{GGSZ,BPGAMMA}.
A data set of $0.8\,{\rm fb^{-1}}$  is analysed and the parameters extracted from measuring the relative rates of various event classes, for example events containing both a neutral \PD decay to the signal mode and a neutral \PD decay to an even or odd \CP eigenstate.
Binning schemes are defined following the guidance of amplitude models developed at the $B$-factories with the goal of providing a partitioning which provides good statistical sensitivity in the $\gamma/\phi_3$ analysis.
The use of an amplitude model to define the binning introduces no model-dependent bias in the measurement of the \ci and \si parameters.  However if the model does not describe nature well then the statistical power of the binning choice will be diluted with respect to expectation. 

In $\PDz\to\KsPiPi$ decays the coefficients are measured in 8 pairs of Dalitz plane bins.   Results are given for four choices of binning: two where the Dalitz space is partitioned into equal intervals of the strong-phase difference, with the variation of the strong-phase difference assumed to follow the models developed by either BaBar~\cite{BaBarCISI} (`equal $\Delta \delta_D$ BaBar') or Belle~\cite{BELLECISI}  (`equal $\Delta \delta_D$ Belle'); and two where the binning is optimised to yield the best possible sensitivity to $\gamma/\phi_3$  in a low background (`optimal binning') and higher background (`modified optimal binning') environment. 
For the equal interval strong-phase analysis the typical uncertainties on each \ci (\si) measurement range from around $0.06$ to $0.16$ ($0.11$ to $0.21$), depending on the bin.   

In $\PDz\to\KsKK$ decays the \ci and \si coefficients are measured in an equal interval strong-phase binning, derived from the model presented in~\cite{BaBarKSKKCISI}.  The lower statistics for this decay restrict the analysis to a fewer number of bins than is the case for $\PDz\to\KsPiPi$.  Results are presented for the Dalitz space partitioned into pairs of 2, 3 and 4 bins.  With 4 pairs of bins the precision on \ci varies between $0.07$ and $0.36$ for \ci, and $0.31$ and $0.73$ for \si.

\begin{figure}
\centering
\includegraphics[width=0.48\textwidth]{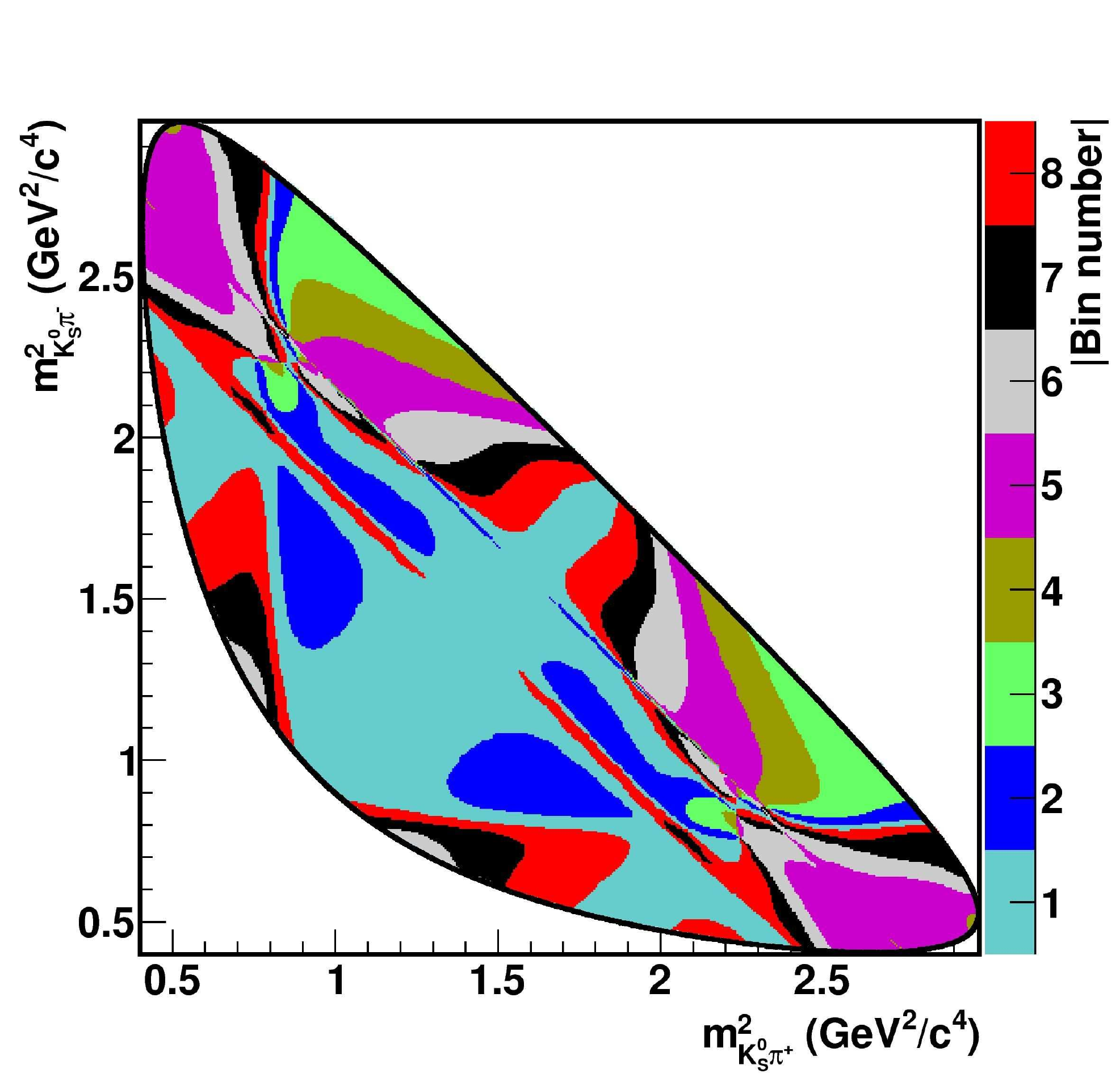}
\includegraphics[width=0.48\textwidth]{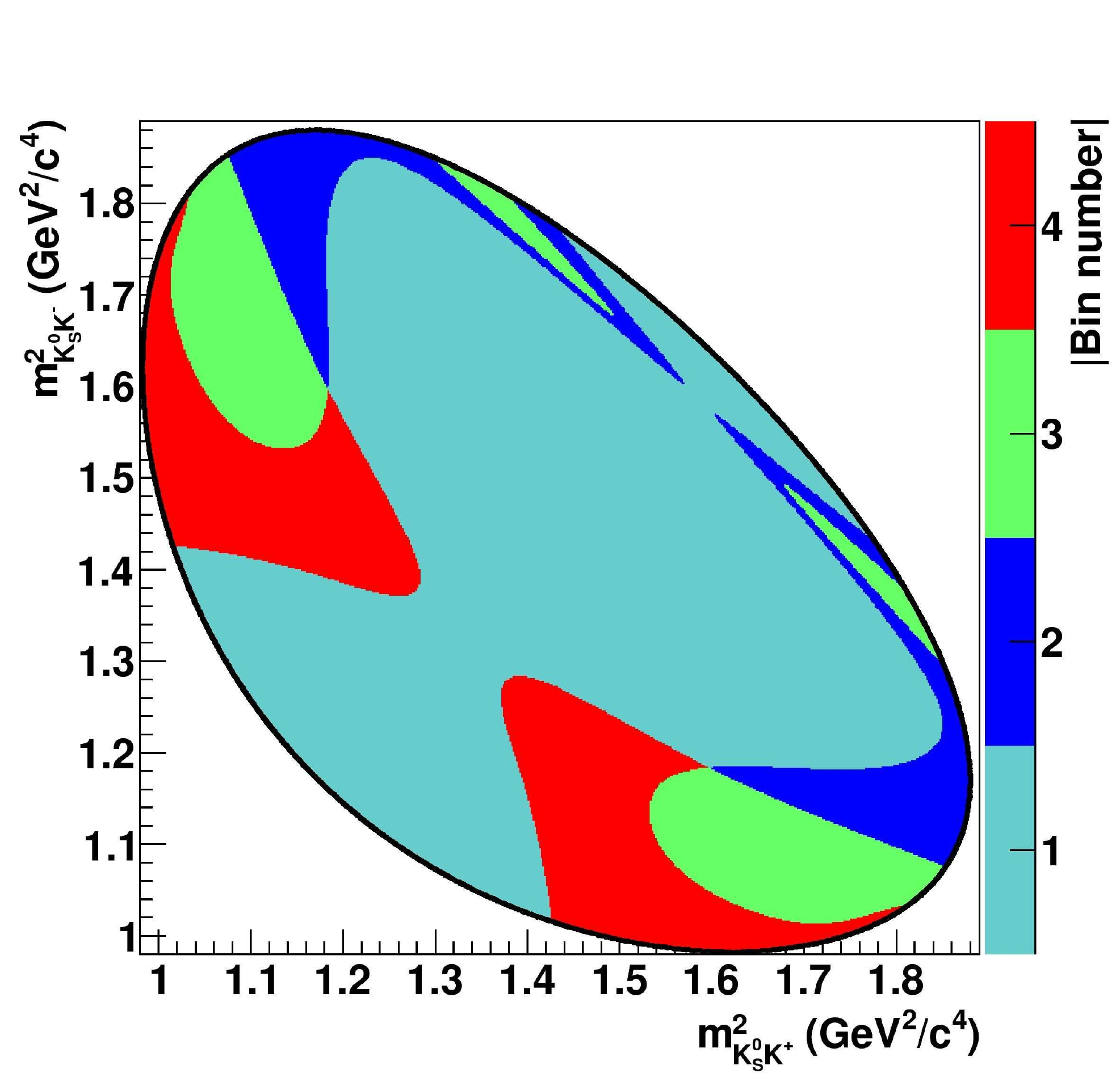}
\caption{Equal  $\Delta \delta_D$ BaBar binning for \KsPiPi (left) and for \KsKK (right).}
\label{fig:binplots}
\end{figure}

The bin definitions for the equal $\Delta \delta_D$ BaBar \KsPiPi scheme and that for \KsKK with 4 pairs of bins is shown in Fig.~\ref{fig:binplots}.

The knowledge of \ci and \si for both decays can be improved by future measurements.  Already the BES~III experiment has accumulated around $3~{\rm fb^{-1}}$ at the $\psi(3770)$ resonance and a total integrated luminosity of  $5-10~{\rm fb^{-1}}$ is foreseen at this collision energy~\cite{BESIII}.  Future facilities, such as a $\tau$-charm factory in Novosibirsk~\cite{NOVOSIBIRSK} or the SuperB project operating at the charm threshold~\cite{SUPERB} would provide substantially larger samples of up  to $500~{\rm fb^{-1}}$.
% SuperB claims 500 fb-1

\section{Fit studies}

In order to determine the projected uncertainties on the mixing and \CP violation parameters in the model-independent approach a programme of `toy Monte Carlo' studies is conducted.  In each study 
events are generated in bins of Dalitz space and \PDz decay time according to Eqn.~\ref{eq:KpTBinCPV}.
Unless stated otherwise, the binning scheme adopted  is the equal $\Delta \delta_D$ BaBar binning with 8 pairs of bins for \KsPiPi and the 4 pairs of bins for \KsKK. 
For the results reported below \xD and \yD and the \CP violation parameters are set to their current world average values~\cite{HFAG}.  The number of events in each bin is smeared to account for statistical fluctuations.  No background or detector effects are included.  A binned maximum likelihood fit is then applied to determine the parameters of interest.  Many samples are generated and fit, and in the case of successful convergence the fit parameters are recorded for each simulated experiment.  The precision on each parameter of interest is determined by taking the standard deviation of the set of fit values.

In general the conclusions of each study have a dependence on the sample size.  Therefore three scenarios are considered:
\begin{enumerate}
\item{Data sets of a similar magnitude to those that have already been analysed at the $B$-factory experiments.  In the case of \KsPiPi this is 0.5M decays, and for \KsKK 0.1M decays.}
\item{Data sets that may become available at LHCb after several years of operation.  Sample sizes of 10M and 2M are assumed for \KsPiPi and \KsKK respectively.}
\item{Data sets that may become available at an upgraded LHCb or a future high luminosity $e^+e^-$ experiment such as Belle-II or SuperB.  Sample sizes of 100M and 20M are assumed for \KsPiPi and \KsKK respectively.}
\end{enumerate}
In the case of 2 and 3 the numbers are indicative, and should not be interpreted as official estimates for any of the experiments mentioned.

\subsection{Fitting with external input for ($\boldsymbol{{c_i,\ s_i}}$)}
\label{sec:cisiinput}

In the first set of  studies the \ci and \si parameters, as measured by CLEO-c, are taken as external input, and the mixing and \CP parameters \xD, \yD, \rCP and \aCP are determined from the fit.    The results for 100M \KsPiPi decays are presented in Table~\ref{tab:100Mfitvalues}. 

In one set of experiments, denoted `(\ci, \si) fixed',  the values of \ci and \si are fixed to be the CLEO-c central values in both the generation and fit;  in another, denoted `(\ci, \si) smeared', the parameters used in the generation are smeared in a correlated manner according to their measurement uncertainties, but kept fixed to their central values in the fit.
Comparison between these two scenarios allows the impact of the (\ci, \si) measurement uncertainties on the results for the mixing and \CP violation parameters to be assessed. 

It is also possible to use external input for the \Ti parameters.
Results are available from CLEO-c~\cite{CLEOCISI}, but significantly more precise measurements can be made with the larger flavour tagged samples that are already available from other experiments.
In the studies three scenarios are considered: fixing the values of \Ti in the generation and the fit to those measured by CLEO-c (`\Ti fixed'); 
smearing the values in the generation according to the measurement uncertainties that would be achievable with existing $B$-factory data sets~\cite{ MIXKSPP_BELLE}, but keeping them fixed to their central values in the fit (`\Ti smeared'); fixing the values of \Ti in the generation but including them as free parameters in the fit (`\Ti floating').

\begin{table}[htp]
\centering
\caption{Expected uncertainties on mixing and \CP parameters for 100M \KsPiPi decays, in a variety of fit scenarios.
The values in parentheses indicate the precision on each number. \vspace*{0.15cm}}
\begin{tabular}{l|l l l l}
\hline\hline
Scenario \T\B & $\sigma(\xD)$ (\%) & $\sigma(\yD)$ (\%) & $\sigma(\rCP)$ & $\sigma(\aCP)$ (\dg) \\
\hline
(\ci, \si) fixed,   \Ti fixed     \T & $ 0.0124(2)$  & $ 0.0135(2)$  & $ 0.0116(2)$ & $ 0.740(10)$ \\
(\ci, \si) fixed,   \Ti smeared      & $ 0.2168(29)$ & $ 0.2424(32)$ & $ 0.0497(7)$ & $ 3.721(49)$ \\
(\ci, \si) fixed,   \Ti floating     & $ 0.0173(2)$  & $ 0.0189(2)$  & $ 0.0118(2)$ & $ 0.744(10)$ \\
(\ci, \si) smeared, \Ti floating  \B & $ 0.0757(10)$ & $ 0.0886(11)$ & $ 0.0262(3)$ & $ 1.188(15)$ \\
\hline\hline
\end{tabular}
\label{tab:100Mfitvalues}
\end{table}

For sample sizes of a few million events and above it is found that the uncertainties on the current measurements of \Ti limit the precision with which 
the mixing and \CP parameters can be determined.  An estimate of the systematic uncertainty on each parameter coming from this
source can be obtained by subtracting in quadrature the results for `\Ti fixed' from those of `\Ti smeared'. This exercise yields errors of approximately 
%Quadrature subtract: (\ci, \si) fixed, \Ti smeared (-) (\ci, \si) fixed, \Ti fixed
$0.216(3) \%$ for \xD, $0.242(3) \%$ for \yD, $0.0483(7)$ for \rCP and $3.65(5)\dg$ for \aCP. 
Improved precision is obtained if the \Ti values are left as free parameters in the fit.   
This approach is therefore considered as baseline for the subsequent discussion. 

The statistical uncertainties from the \KsPiPi analysis, as estimated from the `(\ci, \si) fixed, \Ti floating'
ensemble of experiments, for the three sample sizes under consideration are shown in Table~\ref{tab:kspipisum}.  In addition are listed 
the estimated systematic uncertainties in the binned analysis arising from the current
knowledge of the (\ci, \si) parameters.  These errors are determined by subtracting in quadrature the mean uncertainties of  the `(\ci, \si) fixed, \Ti fixed'  event experiments from those of  the corresponding `(\ci, \si) smeared, \Ti fixed' studies.   
Also shown are the statistical and model uncertainties of an unbinned model-dependent Belle analysis \cite{MIXKSPP_BELLE} conducted with around 0.5M decays.

\begin{table}[htp]
\centering
\caption{Expected statistical and systematic uncertainties from knowledge of (\ci, \si) in a binned \KsPiPi analysis for a variety of sample sizes.
The values in parentheses indicate the precision on each number.  Also shown are the statistical and model-related systematic uncertainties
from an unbinned Belle analysis of 0.5M signal events~\cite{MIXKSPP_BELLE}. \vspace*{0.15cm}}
\begin{tabular}{l|c c c c|c c}
\hline\hline
\T\B & \multicolumn{4}{c|}{Binned} & \multicolumn{2}{c}{Unbinned, 0.5M} \\
\cline{2-7}
Parameter \T\B & Stat, 0.5M & Stat, 10M & Stat, 100M & Syst (\ci, \si) & Stat & Syst model \\ 
\hline
$\sigma(\xD)$ (\%)   \T & $0.251(3)$  & $0.054(1)$ & $0.017(0)$ & $0.076(1)$ & $0.30$             & $^{+0.09}_{-0.16}$ \\
$\sigma(\yD)$ (\%)      & $0.272(4)$  & $0.061(1)$ & $0.019(0)$ & $0.087(1)$ & $0.25$             & $^{+0.07}_{-0.08}$ \\
$\sigma(\rCP)$          & $0.175(2)$  & $0.037(1)$ & $0.012(0)$ & $0.024(0)$ & $^{+0.30}_{-0.29}$ & $0.08$             \\
$\sigma(\aCP)$ (\dg) \B & $12.46(16)$ & $2.42(3)$  & $0.74(1)$  & $0.88(2)$  & $^{+16}_{-18}$     & $^{+2}_{-4}$       \\
\hline\hline
\end{tabular}
\label{tab:kspipisum}
\end{table}
%first three columns are row labelled (\ci, \si) fixed, \Ti floating. order = 100M, 10M, 0.5M
%(\ci, \si) fixed, \Ti floating & $ 0.0173 \pm 0.0002$ & $ 0.0189 \pm 0.0002$ & $ 0.0118 \pm 0.0002$ & $ 0.744 \pm 0.010$ \\
%(\ci, \si) fixed, \Ti floating & $ 0.0541 \pm 0.0007$ & $ 0.0607 \pm 0.0008$ & $ 0.0368 \pm 0.0005$ & $ 2.417 \pm 0.032$ \\
%(\ci, \si) fixed, \Ti floating & $ 0.2511 \pm 0.0033$ & $ 0.2715 \pm 0.0036$ & $ 0.1751 \pm 0.0023$ & $ 12.460 \pm 0.164$ \\
%syst (ci, si) column is ((\ci, \si) smeared, \Ti fixed) (-) ((\ci, \si) fixed, \Ti fixed) for 100M events
%(\ci, \si) smeared, \Ti fixed & $ 0.0772 \pm 0.0010$ & $ 0.0876 \pm 0.0012$ & $ 0.0264 \pm 0.0003$ & $ 1.150 \pm 0.015$ \\
%(\ci, \si) fixed,   \Ti fixed & $ 0.0124 \pm 0.0002$ & $ 0.0135 \pm 0.0002$ & $ 0.0116 \pm 0.0002$ & $ 0.740 \pm 0.010$ \\

The results in Table~\ref{tab:kspipisum} indicate that the statistical uncertainties 
in the binned analysis exhibit an approximate $1/\sqrt{N}$ scaling, as expected.
The 0.5M binned statistical uncertainties are similar to those of the unbinned Belle analysis.
Indeed, in some cases the binned uncertainties are slightly smaller than those from the unbinned study.  It is
assumed that this behaviour can be attributed to
the idealised nature of the toy Monte Carlo and the particular properties of the Belle data set.
In principle the small loss of  information  that occurs in a binned study should lead to a slightly 
reduced statistical precision with respect to an unbinned analysis.  Further discussion on this
issue can be found in Sect.~\ref{sec:binning}.

Comparison of the (\ci, \si) systematic uncertainties with the statistical uncertainties associated with a data set of a 
given size allows  the potential of the available CLEO-c measurements in the \KsPiPi mixing and \CP violation analysis to be assessed.
Of most interest are the \CP violation parameters \rCP and \aCP.  Here the (\ci, \si) uncertainties are smaller than
the statistical uncertainties for a 10M event data set. By interpolating the statistical precision it is found
that the (\ci, \si) uncertainty becomes dominant in data sets of around 
%Reasoning: 10M * (0.037/0.024)^2 ~ 25M
25M and 
%Reasoning: 10M * (2.42/0.88)^2 ~ 75M
75M 
events, for \rCP and \aCP respectively.
Even in the case of 100M events it would be presumably be possible to make the (\ci, \si) uncertainty subdominant, merely by
repeating the CLEO-c analysis with the 3~{$\rm fb^{-1}$} $\psi(3770)$ sample that BES-III now possesses. 
%Reasoning: sqrt(3/0.818) = 1.9 ; 0.024/1.9 = 0.013 ; 0.88/1.9 = 0.46
The current (\ci, \si) uncertainties are significantly smaller than the assigned model errors in the Belle analysis.  
Although larger data sets will allow for the \KsPiPi model to be refined, it is difficult to assess what improvements can be expected in the associated systematic uncertainties.

The precision on the mixing parameters \xD and \yD more rapidly becomes limited by the current knowledge of (\ci, \si).  For 10M events these strong-phase uncertainties are dominant.  They are of a similar size to the assigned model uncertainties in the unbinned Belle analysis.  
Analysis of the quantum-correlated \PD--\PDb data set already available at BES-III
would allow for the (\ci, \si)  uncertainty to remain subdominant for \KsPiPi sample sizes up to around 20M events. 
%Reasoning: 
% sqrt(3/0.818) = 1.9 ; 0.076/1.9 = 0.040 ; 0.087/1.9 = 0.045 ; 10M * (0.054/0.040)^2 = 18M ; 10M * (0.061/0.045) = 18M

\begin{table}[htp]
\centering
\caption{Expected statistical and systematic uncertainties from knowledge of (\ci, \si) in a binned \KsKK analysis for a variety of sample sizes.
The values in parentheses indicate the precision on each number. \vspace*{0.15cm} }
\begin{tabular}{l |c c c c}
\hline\hline
Parameter \T\B & Stat, 0.1M & Stat, 2M  &  Stat, 20M & Syst (\ci, \si)   \\ 
\hline
$\sigma(\xD)$ (\%)    \T & $0.807(11)$ & $0.168(2)$ & $0.066(1)$ & $0.330(4)$ \\
$\sigma(\yD)$ (\%)       & $0.546(7)$  & $0.102(1)$ & $0.033(0)$ & $0.178(2)$ \\
$\sigma(\rCP)$           & $0.266(4)$  & $0.073(1)$ & $0.029(0)$ & $0.068(1)$ \\
$\sigma(\aCP)$ (\dg)  \B & $33.55(44)$   & $6.47(9)$  & $2.34(3)$  & $3.88(7)$ \\
\hline\hline
\end{tabular}
\label{tab:kskksum}
\end{table}
%first three columns are row labelled (\ci, \si) fixed, \Ti floating. order = 20M, 2M, 0.1M
%(\ci, \si) fixed, \Ti floating & $ 0.0663 \pm 0.0009$ & $ 0.0327 \pm 0.0004$ & $ 0.0294 \pm 0.0004$ & $ 2.341 \pm 0.031$ \\
%(\ci, \si) fixed, \Ti floating & $ 0.1679 \pm 0.0022$ & $ 0.1020 \pm 0.0013$ & $ 0.0733 \pm 0.0010$ & $ 6.473 \pm 0.085$ \\
%(\ci, \si) fixed, \Ti floating & $ 0.8071 \pm 0.0106$ & $ 0.5459 \pm 0.0072$ & $ 0.2662 \pm 0.0035$ & $ 33.548 \pm 0.442$ \\
%syst (ci, si) column is ((\ci, \si) smeared, \Ti fixed) (-) ((\ci, \si) fixed, \Ti fixed) for 100M events
%(\ci, \si) smeared, \Ti fixed & $ 0.3317 \pm 0.0044$ & $ 0.1791 \pm 0.0024$ & $ 0.0718 \pm 0.0009$ & $ 4.398 \pm 0.058$ \\
%(\ci, \si) fixed,   \Ti fixed & $ 0.0368 \pm 0.0005$ & $ 0.0225 \pm 0.0003$ & $ 0.0227 \pm 0.0003$ & $ 2.073 \pm 0.027$ \\

The statistical uncertainties that are expected for a binned analysis of \KsKK data, assuming the four-bin equal strong-phase binning, are shown in Table~\ref{tab:kskksum}. 
As expected the precision is worse than in the \KsPiPi case.
The uncertainties on the mixing parameters for the 0.1M scenario are similar to the values of  $0.92\%$ for \xD and $0.57\%$ for \yD that are obtained in an unbinned BaBar analysis\cite{MIXKSPP_BaBar}.    
The (\ci, \si) uncertainties from the CLEO-c analysis are dominant in the measurement of \xD and \yD for the 2M sample,  but not so for \rCP and \aCP.  For 20M events the (\ci, \si) uncertainties on \rCP and \aCP are around twice the size of the expected statistical uncertainties.  Exploitation of the already available BES-III $\psi(3770)$ data would allow for the (\ci, \si) uncertainties to be approximately halved.
%Reasoning: sqrt(3/0.818) = 1.9

\subsection{Fitting with external input for $\boldsymbol{x_D}$ and $\boldsymbol{y_D}$}

An alternative strategy to that presented above is to consider \rCP and \aCP as the principal parameters of interest, and to use external measurements to fix \xD and \yD in the fit.   In this approach it is natural also to determine (\ci, \si) from the fit, as in most scenarios this may be done with a precision that is similar to or better than that of the CLEO-c analysis.   Improved knowledge of (\ci, \si) is useful for the measurement of $\gamma/\phi_3$ using $B^- \to D( K^0_{\rm S}h^+h^-) K^-$ decays.

The free parameters are therefore  \rCP, \aCP, (\ci, \si), \Ti, $\Gamma$  and the overall number of events.  The values of \xD and \yD are fixed to their current world-average values~\cite{HFAG}
% of $\xD = 0.63^{+0.19}_{-0.20}$ and $\yD = 0.75 \pm 0.12$  
in both the generation and the fit.  The statistical uncertainties on \rCP and \aCP and the systematic uncertainties arising from the current knowledge of the mixing parameters, given separately for \xD and \yD, are shown in Table~\ref{tab:kspipi_fixedxdyd}.  The statistical precision is somewhat worse than that obtained  following the procedure presented in Sect.~\ref{sec:cisiinput}.  The systematic uncertainty from the current knowledge of (\xD, \yD) does not dominate the measurement of \rCP until the sample size exceeds 10M decays, with the error associated with \xD being the most significant. For \aCP the measurement remains statistics-limited up to even larger data sets, making this an attractive and complementary alternative analysis strategy to that discussed in Sect.~\ref{sec:cisiinput}.

\begin{table}[htp]
\centering
\caption{Expected statistical and systematic uncertainties on the \CP violation parameters from knowledge of  \xD and \yD in a binned \KsPiPi analysis for a variety of sample sizes.
The values in parentheses indicate the precision on each number. \vspace*{0.15cm} }
\begin{tabular}{l |c c c c c}
\hline\hline
Parameter \T\B & Stat, 0.5M & Stat, 10M  &  Stat, 100M & Syst (\xD) & Syst (\yD)   \\
\hline
$\sigma(\rCP)$       \T & $0.3129(44)$ & $0.0561(8)$ & $0.0170(2)$ & $0.0414(7)$ & $0.0234(5)$ \\
$\sigma(\aCP)$ (\dg) \B & $27.26(38)$  & $3.35(5)$   & $0.99(1)$ & $0.51(4)$ & $0.29(7)$ \\
\hline\hline
\end{tabular}
\label{tab:kspipi_fixedxdyd}
\end{table}
%search for (\ci, \si) floating, \Ti floating. order is 100M, 10M, 0.5M
%(\ci, \si) floating, \Ti floating & $ 0.0170 \pm 0.0002$ & $ 0.989 \pm 0.013$ \\
%(\ci, \si) floating, \Ti floating & $ 0.0561 \pm 0.0008$ & $ 3.354 \pm 0.048$ \\
%(\ci, \si) floating, \Ti floating & $ 0.3129 \pm 0.0044$ & $ 27.259 \pm 0.381$ \\
%search for (\ci, \si) floating, \Ti floating given smear x by WA for 100M and subtract in quadrature
%(\ci, \si) floating, \Ti floating & $ 0.0448 \pm 0.0006$ & $ 1.113 \pm 0.015$ \\
%search for (\ci, \si) floating, \Ti floating given smear y by WA for 100M and subtract in quadrature
%(\ci, \si) floating, \Ti floating & $ 0.0289 \pm 0.0004$ & $ 1.031 \pm 0.014$ \\

The uncertainties on \ci and \si from a sample of 100M \KsPiPi events are presented in Table~\ref{tab:FitSixySyst}.  The results depend on how well the input parameters \xD and \yD are known, and so the study is repeated considering a variety of scenarios in which the uncertainty on both parameters is scaled from their current values down by a common factor.   Also shown are the measurement uncertainties from the CLEO-c quantum-correlated analysis.   It can be seen that even without any  improvement in the knowledge of the mixing parameters it is possible to fit \ci and \si in certain bins with a precision that is similar or better than that of the CLEO-c measurements.   If the knowledge of the mixing parameters improves by a factor of three or four in precision, for example through mixing analyses using alternative decay modes or through analysis of independent data sets,  then it is possible to reduce significantly the uncertainties on the (\ci, \si) parameters.     

\begin{table}[htp]
\centering
\caption{Fit uncertainties on  \ci and \si for 100M \KsPiPi decays when fixing (\xD, \yD) and assuming a certain knowledge of 
these parameters with respect to the current precision.   Also shown are the current measurement uncertainties on \ci and \si from CLEO-c~\cite{CLEOCISI}. \vspace*{0.15cm}}
\begin{tabular}{c|c c c c c|c}
\hline\hline
\T\B           & \multicolumn{5}{c|}{Scale factor for current uncertainties on \xD and \yD} & \\
\cline{2-6}
\T\B Parameter & 1 & $1/2$ & $1/3$ & $1/4$ & $1/\infty$ & CLEO-c \\
\hline
$\sigma(c_1)$ \T & $0.125(2)$ & $0.087(1)$ & $0.079(1)$ & $0.076(1)$ & $0.072(1)$ & 0.055 \\
$\sigma(c_2)$ & $0.085(1)$ & $0.062(1)$ & $0.056(1)$ & $0.053(1)$ & $0.050(1)$ & 0.093 \\
$\sigma(c_3)$ & $0.068(1)$ & $0.064(1)$ & $0.064(1)$ & $0.062(1)$ & $0.064(1)$ & 0.161 \\
$\sigma(c_4)$ & $0.108(2)$ & $0.079(1)$ & $0.071(1)$ & $0.068(1)$ & $0.065(1)$ & 0.153 \\
$\sigma(c_5)$ & $0.159(2)$ & $0.091(1)$ & $0.071(1)$ & $0.060(1)$ & $0.048(1)$ & 0.062 \\
$\sigma(c_6)$ & $0.114(2)$ & $0.083(1)$ & $0.079(1)$ & $0.076(1)$ & $0.070(1)$ & 0.126 \\
$\sigma(c_7)$ & $0.065(1)$ & $0.056(1)$ & $0.056(1)$ & $0.056(1)$ & $0.056(1)$ & 0.163 \\
$\sigma(c_8)$ \B & $0.103(1)$ & $0.071(1)$ & $0.065(1)$ & $0.062(1)$ & $0.058(1)$ & 0.102 \\
\hline
$\sigma(s_1)$ \T & $0.102(1)$ & $0.102(1)$ & $0.100(1)$ & $0.098(1)$ & $0.097(1)$ & 0.107 \\
$\sigma(s_2)$ & $0.123(2)$ & $0.089(1)$ & $0.075(1)$ & $0.067(1)$ & $0.057(1)$ & 0.195 \\
$\sigma(s_3)$ & $0.225(3)$ & $0.154(2)$ & $0.118(2)$ & $0.099(1)$ & $0.076(1)$ & 0.138 \\
$\sigma(s_4)$ & $0.169(2)$ & $0.124(2)$ & $0.099(1)$ & $0.091(1)$ & $0.078(1)$ & 0.214 \\
$\sigma(s_5)$ & $0.055(1)$ & $0.052(1)$ & $0.051(1)$ & $0.051(1)$ & $0.049(1)$ & 0.134 \\
$\sigma(s_6)$ & $0.219(3)$ & $0.151(2)$ & $0.123(2)$ & $0.105(1)$ & $0.088(1)$ & 0.205 \\
$\sigma(s_7)$ & $0.236(3)$ & $0.161(2)$ & $0.119(2)$ & $0.094(1)$ & $0.065(1)$ & 0.154 \\
$\sigma(s_8)$ \B & $0.170(2)$ & $0.117(2)$ & $0.096(1)$ & $0.084(1)$ & $0.070(1)$ & 0.158 \\
\hline\hline
\end{tabular}
\label{tab:FitSixySyst}
\end{table}

As can be seen from Table~\ref{tab:FitSixySyst} the statistical precision is better in some bins compared to others,  and the measurement uncertainties from CLEO-c also vary from bin-to-bin. Therefore analysis of sample sizes of $\sim 20\,\mathrm{M}$ decays would already be beneficial for improving knowledge of certain parameters.

A similar analysis performed on \KsKK simulated data yields the same qualitative conclusions.  Again the precision on the \CP violation parameters is inferior to that observed in Sect.~\ref{sec:cisiinput}, but it is possible to measure the $(\ci, \si)$ parameters with a precision that is useful for future $\gamma/\phi_3$ studies.

\subsection{Fitting with no external inputs}

With a sufficiently large data set it is possible to determine the mixing, \CP violation and (\ci, \si) parameters in the fit.
In practice this is achieved through a two-stage procedure.  In the first stage the parameters of small magnitude in Equation~\ref{eq:KpTBinCPV}, namely \xD, \yD, \ci and \si, are fixed to their current world-average values and the other parameters are determined.  
In the second stage the parameters that were fitted in the first stage are fixed, and vice versa.   

When this study is performed for ensembles of 100M \KsPiPi decays uncertainties of  
$0.0644(10)\%$, $0.0790(12)\%$, $0.0244(4)$ and $1.762(27)\dg$ are obtained for \xD, \yD, \rCP and \aCP respectively.  These results are not competitive with those of the strategies that make use of external inputs.
The precision for (\ci, \si) in this study is on average around twice as poor as  that obtained when constraining \xD and \yD to lie within their current measurement uncertainties.

\subsection{Varying the number of bins}
\label{sec:binning}

It is interesting to consider the variation in the precision of the fitted parameters if the model-independent analysis is repeated with different numbers of bins.
As measurements only exist for 8 pairs of bins in the case of \KsPiPi and for 2, 3 and 4 pairs of  bins in the case of \KsKK, the study instead uses input values of (\ci, \si) and \Ti as calculated from the model in Ref.~\cite{BaBarKSKKCISI}, and these are assumed to be known perfectly.  The free parameters of the fit are \xD, \yD, \rCP, \aCP, $\Gamma$ and the total number of decays.     For \KsPiPi experiments are performed with 8, 16, 32 and 50 pairs of equal-phase bins, and for \KsKK studies are made with 4, 8, 16 and 32 pairs of bins.

\begin{figure}
\centering
\begin{minipage}[t]{0.48\textwidth}
\includegraphics[width=\textwidth]{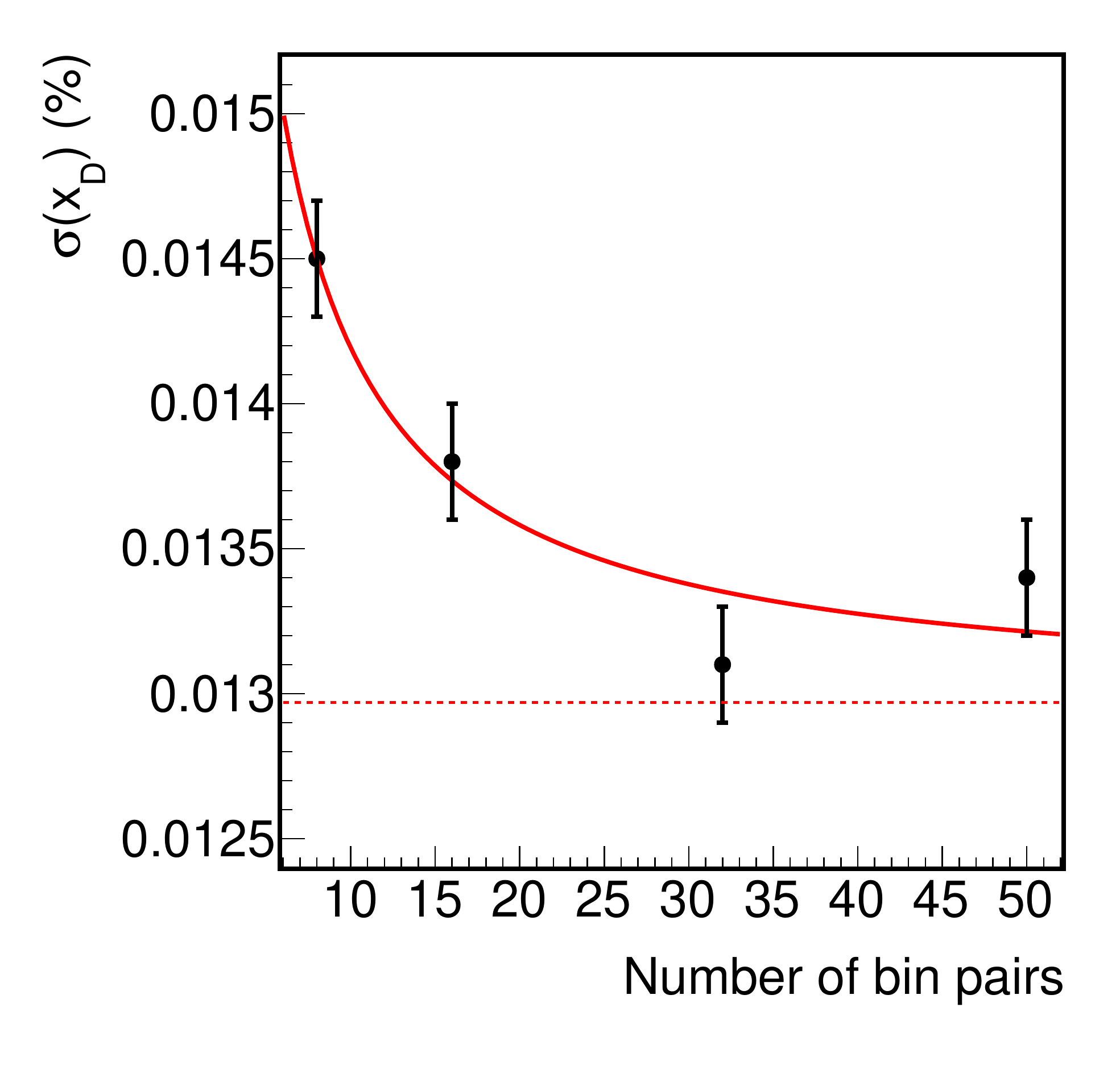}
\end{minipage}
\begin{minipage}[t]{0.48\textwidth}
\includegraphics[width=\textwidth]{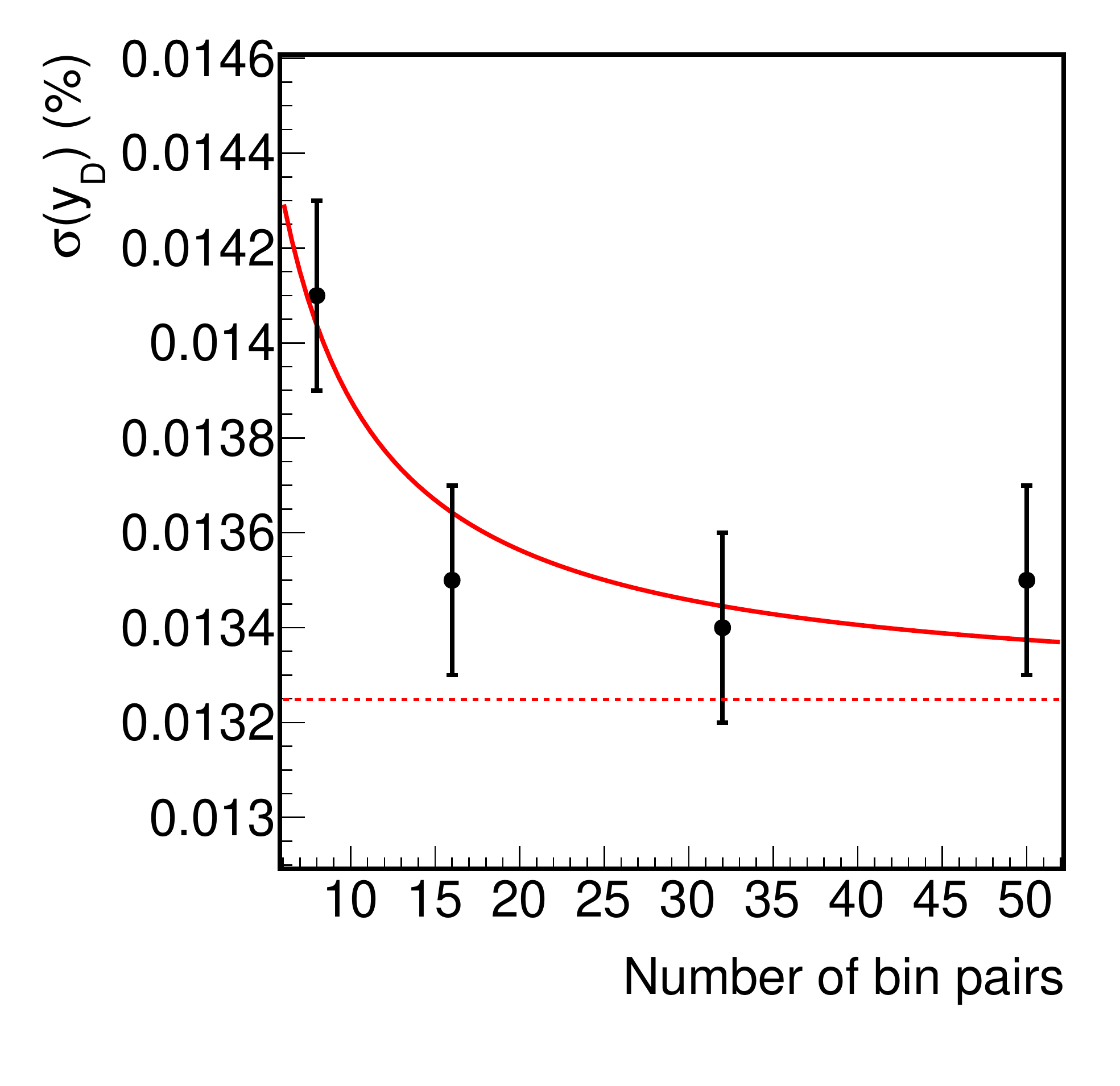}
\end{minipage}
\begin{minipage}[t]{0.48\textwidth}
\includegraphics[width=\textwidth]{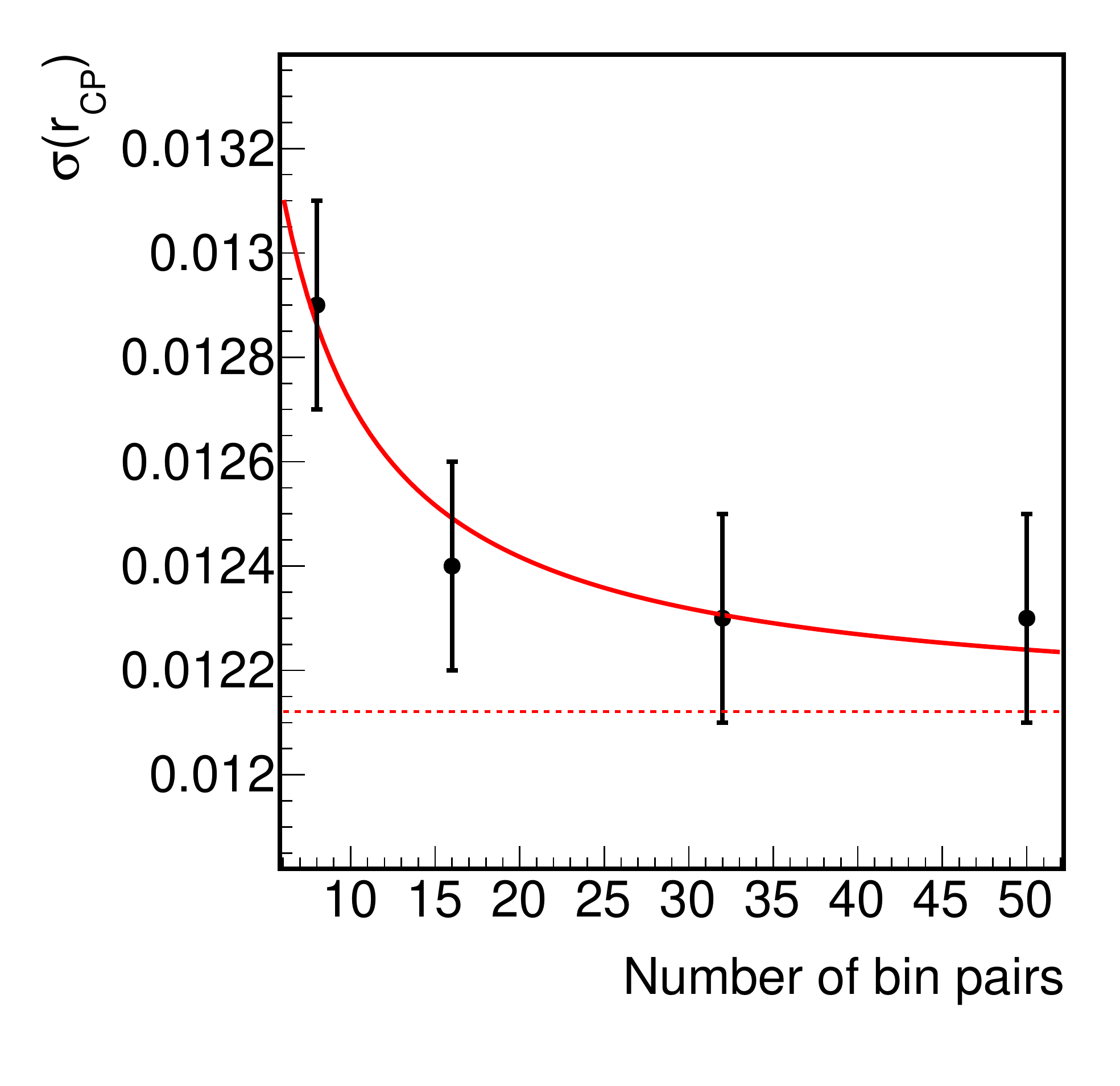}
\end{minipage}
\begin{minipage}[t]{0.48\textwidth}
\includegraphics[width=\textwidth]{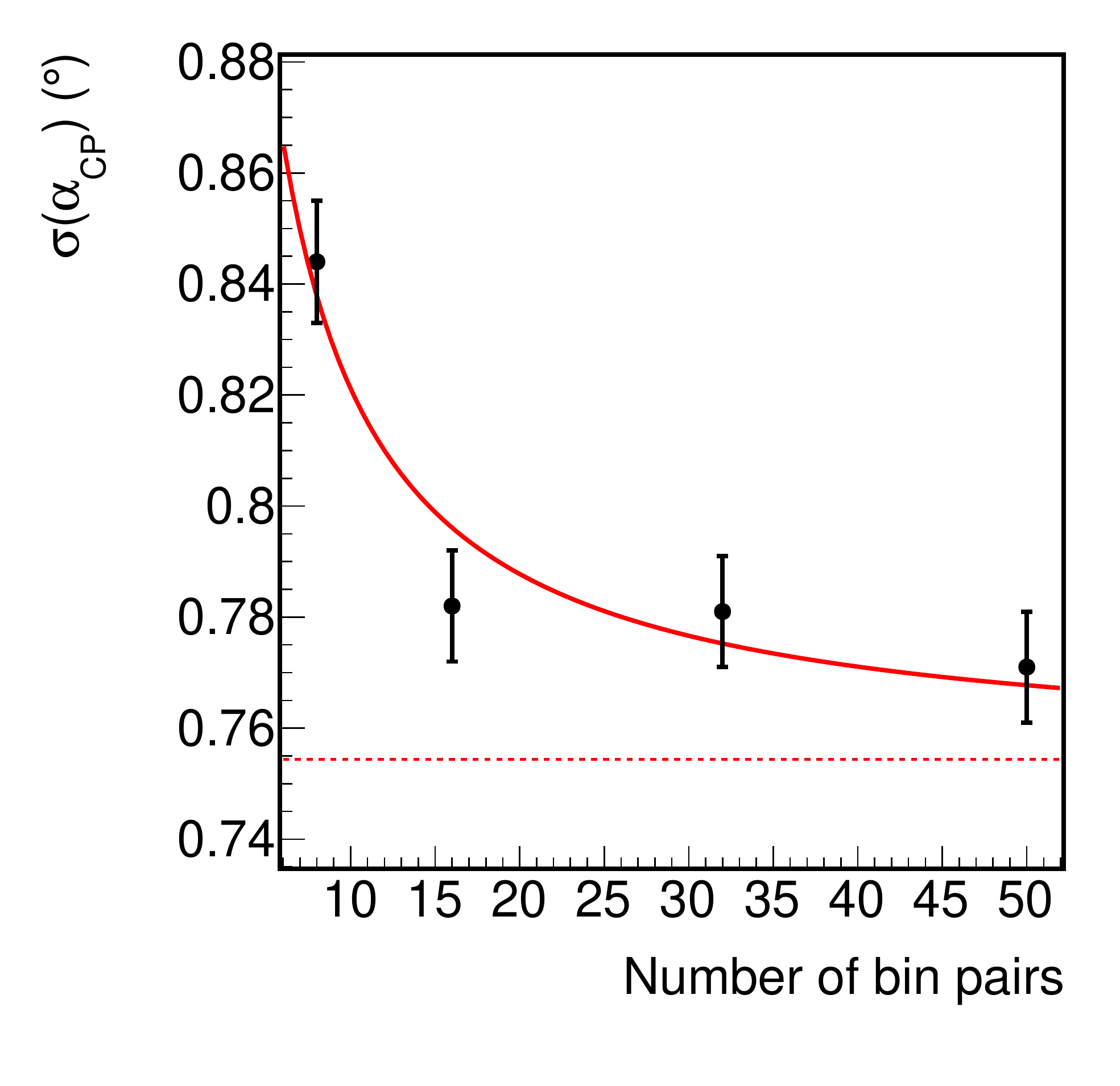}
\end{minipage}
\caption{Dependence of the fit uncertainty for the mixing and \CP violation parameters against number of Dalitz bin pairs for 100M \KsPiPi decays.   The input parameters and fit conditions are described in the text. Superimposed is a fit to Eqn.~\ref{eq:asymptote} (solid line) and the asymptotic value of the fit (dotted). }
\label{fig:binvariationKsPiPi}
\end{figure}

\begin{figure}
\centering
\begin{minipage}[t]{0.48\textwidth}
\includegraphics[width=\textwidth]{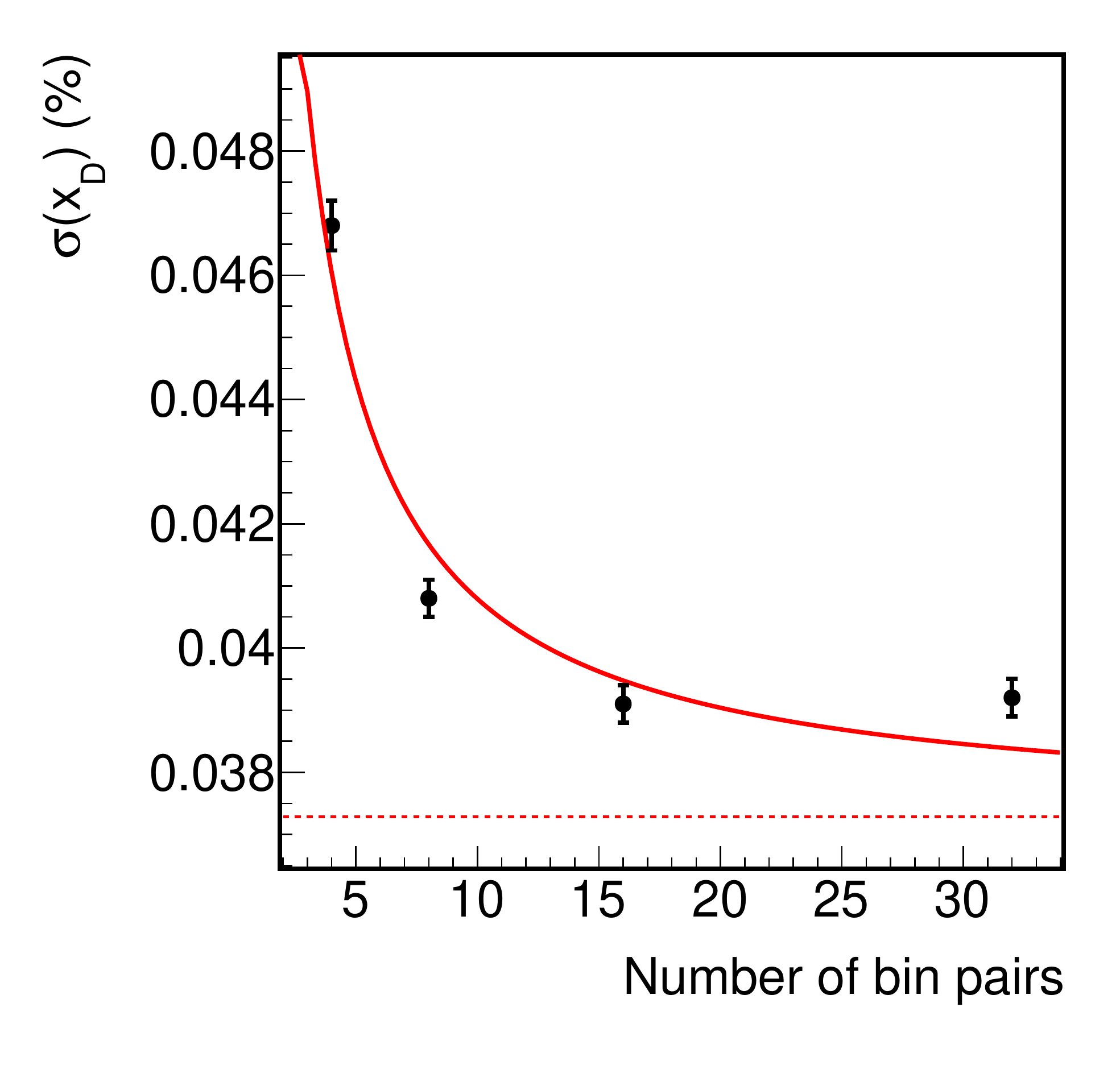}
\end{minipage}
\begin{minipage}[t]{0.48\textwidth}
\includegraphics[width=\textwidth]{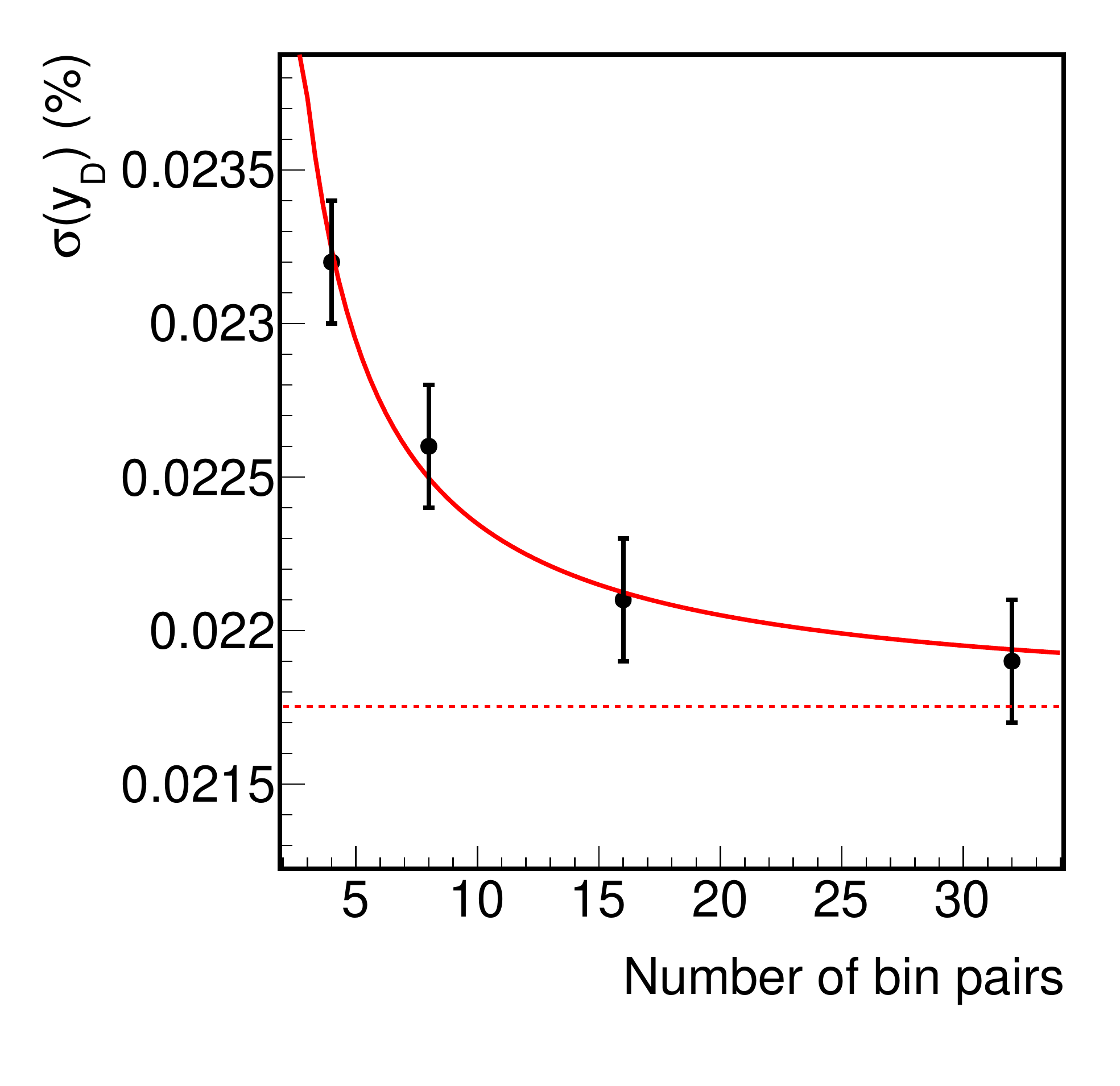}
\end{minipage}
\begin{minipage}[t]{0.48\textwidth}
\includegraphics[width=\textwidth]{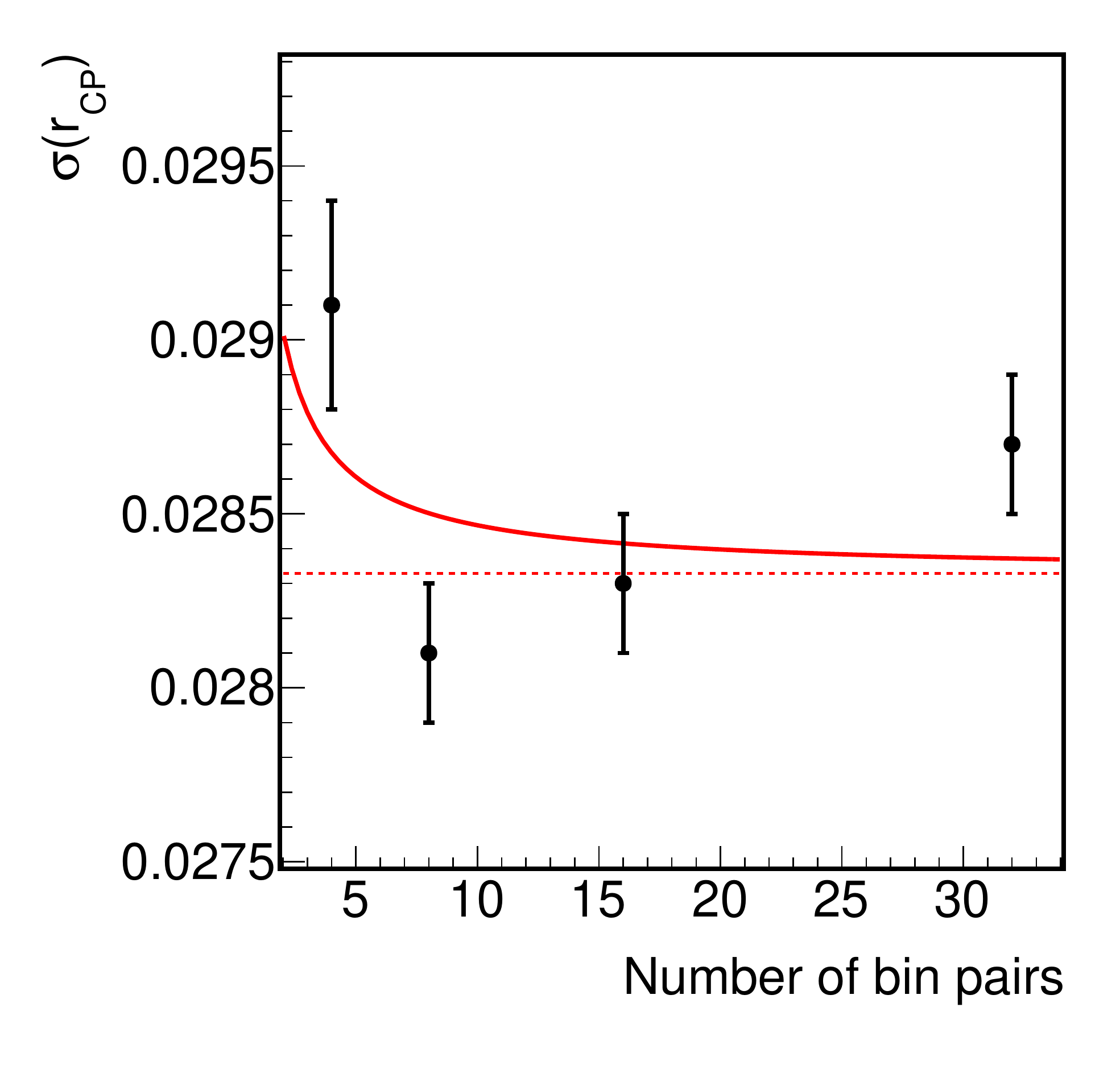}
\end{minipage}
\begin{minipage}[t]{0.48\textwidth}
\includegraphics[width=\textwidth]{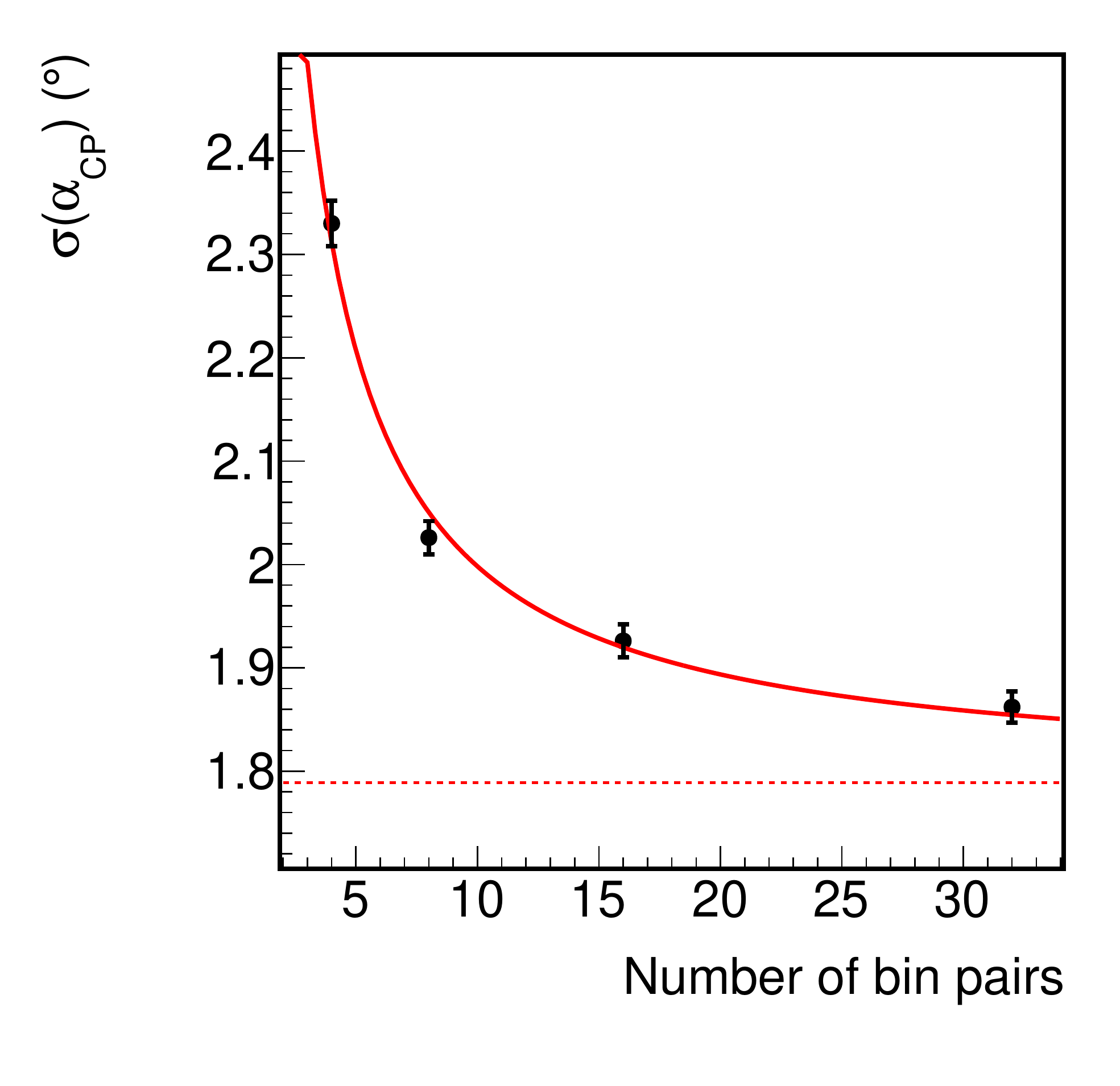}
\end{minipage}
\caption{Dependence of the fit uncertainty for the mixing and \CP violation parameters against number of Dalitz bin pairs for 20M \KsKK decays.  The input parameters and fit conditions are described in the text. Superimposed is a fit to Eqn.~\ref{eq:asymptote} (solid line) and the asymptotic value of the fit (dotted).  }
\label{fig:binvariationKsKK}
\end{figure}

The  precision determined for each scenario is plotted in Figs.~\ref{fig:binvariationKsPiPi} and~\ref{fig:binvariationKsKK} for 100M \KsPiPi and 20M \KsKK decays.   In general the precision improves with the number of bins.  In order to quantify this trend  the following function has been fit to each variable $v$:
\begin{equation}
\sigma(v) = a + \frac{b}{N}
\label{eq:asymptote}
\end{equation}
where $N$ is the number of Dalitz plot bin pairs.  The asymptotic parameter $a$ gives the statistical precision that is achievable in the unbinned scenario.
The values of $a$ and $b$ for each mixing and \CP violation parameter are shown in Table~\ref{tab:varyNbins}.

\begin{table}[htp]
\centering
\caption{Results for the parameters $a$ and $b$, defined in Eqn.~\ref{eq:asymptote}, that describe the dependence of the precision of the mixing and \CP parameters on the number of Dalitz plot bins.}
\vspace*{0.15cm}
\begin{tabular}{l| c c |c c}
\hline\hline
\T\B                 & \multicolumn{2}{c|}{\KsPiPi} & \multicolumn{2}{c}{\KsKK} \\
\cline{2-5}
Parameter       \T\B & $a$         & $b$           & $a$         & $b$ \\
\hline
$\sigma(\xD) (\%)$   \T & $0.0130(2)$ & $0.0122(25)$  & $0.0373(3)$ & $0.0350(21)$ \\
$\sigma(\yD) (\%)$      & $0.0132(2)$ & $0.0063(25)$  & $0.0218(2)$ & $0.0060(11)$ \\
$\sigma(\rCP)$          & $0.0121(2)$ & $0.0059(25)$  & $0.0283(2)$ & $0.0014(15)$ \\
$\sigma(\aCP) (\dg)$ \B & $0.754(9)$  & $0.668(130)$  & $1.789(14)$ & $2.091(115)$ \\
\hline\hline
\end{tabular}
\label{tab:varyNbins}
\end{table}

Although improved precision is in principle achievable by increasing the number of bins from the values for which measurements are currently available, this improvement is in all cases rather modest, even when going to an infinite number of bins ($\sim 10\%$ for most parameters).  This indicates that the model-independent analysis with 8  pairs of bins for \KsPiPi,  and 4 pairs of bins for \KsKK, has a statistical precision which is only marginally inferior to that of the model-dependent, unbinned analysis.  Furthermore there can be no significant improvement in sensitivity by changing the binning scheme in the analysis from the equal-interval strong-phase  binning which has been assumed for the studies presented here.
 In practice the gain in precision when increasing the number of bins will presumably be even  more modest than Figs.~\ref{fig:binvariationKsPiPi} and~\ref{fig:binvariationKsKK} suggest, as the systematic uncertainties associated with the knowledge of (\ci, \si), not included in this study, will grow as the bin sizes become smaller.

\section{Discussion and summary}
\label{sec:conclude}

As already demonstrated by existing measurements, a time-dependent analysis of \KsHH decays is a powerful method for studying 
 mixing and to probe for  indirect \CP violation in the \PDz--\PDzb system.  Observation of \CP violation in charm mixing related observables at a level beyond the very small value predicted in the Standard Model would be a strong indication of New Physics contributions.  A binned, model-independent fit is an interesting alternative analysis strategy to the unbinned model-dependent approach which has so far been pursued.   An extensive programme of simulation studies has been performed to assess the strengths and limitations of the binned analysis.  From these studies three main conclusions emerge:

\begin{enumerate}
\item{A binned  analysis is a very attractive approach for extracting mixing and \CP violation information from  \KsHH
decays. By making use of the measurements of the (\ci, \si) parameters available from CLEO-c it is possible to obtain a statistical precision on \xD, \yD, \rCP and \aCP that is only marginally inferior to that available from an unbinned analysis. Most importantly, the binned analysis has no systematic error arising from the use of an amplitude model, in contrast to the unbinned approach where this source of uncertainty is unavoidable.}
\item{The existing CLEO-c measurements of (\ci, \si), performed with only 0.8~${\rm fb^{-1}}$ of $\psi(3770)$ data are adequate for analysing the size of sample that can be expected at the current LHCb experiment.   Looking further forward to the extremely large samples that may be accumulated at the LHCb upgrade or Super-$B$ or Belle-II it is very unlikely that the uncertainties on \ci and \si will ever become limiting in the measurement of \rCP and \aCP, which are the principal parameters of interest.  Even with 100M \KsPiPi decays the expected statistical uncertainty on the \CP parameters is only around half that of the systematic arising from the CLEO-c measurement errors on \ci and \si.   BES-III already has a $\psi(3770)$ sample that is four times larger than that analysed by CLEO-c, and this is set to grow.     Analysis of these data will allow for the (\ci, \si) systematic uncertainty to become sub-dominant, even for enormous samples of $K^0_{\rm S} h^+h^-$ decays.   Plans exist for future projects, such as a Super-$B$ threshold run, or a Novorsibirsk $\tau-$charm factory, that will provide a  $\psi(3770)$ data set one-to-two orders of magnitude larger than that accumulated  at BES-III.  Although welcome for many reasons, these initiatives are not essential for improving the knowledge of the \CP violation parameters from the analysis of flavour-tagged $K^0_{\rm S} h^+h^-$ data.  On the other hand, the (\ci, \si) uncertainties will more rapidly become limiting for the mixing parameters \xD and \yD,  and here larger quantum-correlated \PD--\PDb samples will be useful.

The bin granularity and partitioning schemes for which the CLEO-c measurements have been performed look well suited for  \PDz--\PDzb mixing and \CP violation  studies with all current and future data sets.
}
\item{With very large \KsHH samples an alternative fit strategy becomes possible which will  benefit the measurement of the unitarity triangle angle $\gamma/\phi_3$ using $B^- \to D(K^0_{\rm S} h^+h^-)K^-$ decays.   
The induced systematic uncertainty on $\gamma/\phi_3$ arising from the CLEO-c measurements of (\ci, \si) is estimated to be 2.1\dg for the equal phase binning~\cite{CLEOCISI}.  This uncertainty is  small compared to the anticipated statistical precision of the current LHCb experiment,  but may become limiting at the upgrade, where a measurement error of $\sim 1 \dg$ is expected.  A similar precision is foreseen at the Super-$B$ experiment.  It is therefore desirable to reduce the uncertainties on (\ci, \si) by a factor of two or better with respect to the CLEO-c analysis.

Such an improvement may
 be possible from analysis of the BES-III data alone.  However, a complementary measurement of similar precision can be performed using flavour-tagged  \KsHH data.  In this approach external constraints are placed on \xD and \yD and the fit is used to determine (\ci, \si), as well as \rCP and \aCP.  Assuming 100M \KsPiPi events, and a knowledge of \xD and \yD which is a factor four times better than at present, then the uncertainties on \ci and \si can  be decreased to around 60\%  (on average) of their current values.   This strategy therefore appears promising, although it requires improved knowledge of \xD and \yD which must come from independent data sets and/or mixing analyses based on other decay modes.
}
\end{enumerate}

In conclusion, the model-independent binned fit has many attractive features to recommend it as an analysis strategy for \KsHH  decays, both with the data sets currently available and the very large samples that will be collected over the coming decade.   First results using this approach are eagerly awaited.

\section*{Acknowledgments}

We acknowledge useful discussions with Anton Poluektov and Jim Libby.  We are grateful for support from 
the STFC, United Kingdom.

\end{document}